\documentclass[prd,aps,twocolumn,showpacs,preprintnumbers,amsmath,amssymb]{revtex4}
\usepackage[dvips]{graphicx} 
\usepackage{amsmath} 
\usepackage{amssymb} 
 
\usepackage{graphicx}
\usepackage{dcolumn}
\usepackage{bm}
\def\be{\begin{equation}} 
\def\ee{\end{equation}} 
\def\bea{\begin{eqnarray}} 
\def\eea{\end{eqnarray}} 
\newcommand{\alp}{\alpha'}
\newcommand{\del}{\partial}
\newcommand{\comment}[1]{}
\newcommand{\mbp}{\left(\frac{\mu}{\alpha'\beta \hat p}\right)^2}
 
\begin{document} 
 
 
\date{\today} 
 
\title{Emergence of Fluctuations from a Tachyonic Big Bang}

\author{Robert H. Brandenberger} \email[email: ]{rhb@hep.physics.mcgill.ca}
\author{Andrew R. Frey }\email[email: ]{frey@hep.physics.mcgill.ca}
\author{Sugumi Kanno} \email[email: ]{sugumi@hep.physics.mcgill.ca}

\affiliation{Department of Physics, McGill University, 
Montr\'eal, QC, H3A 2T8, Canada} 
 
 
\begin{abstract} 

It has recently been speculated that the end state of a collapsing
universe is a \textit{tachyonic big crunch}. The time reversal of this
process would be the emergence of an expanding universe from a
tachyonic big bang. In this framework, we study the emergence of
cosmological fluctuations. In particular, we compare the amplitude
of the perturbations at tne end of the tachyon phase with what would
be obtained assuming the usual vacuum initial conditions. We find
that cosmological fluctuations emerge in a thermal state. We comment
on the relation to the \textit{trans-Planckian problem} of inflationary
cosmology.
 
\end{abstract} 
 
\maketitle

\newcommand{\eq}[2]{\begin{equation}\label{#1}{#2}\end{equation}} 
 
\section{Introduction} 

There have recently been several proposals to study the
resolution of cosmological singularities in string theory.
Postulating a tachyon condensate phase in the very
early universe \cite{tachyoncond} is one of these
new avenues. In this paper, we make a first attempt
to connect such an early stringy phase of cosmology with
cosmological observations. Specifically, we study the
origin of cosmological fluctuations in this framework.

If the initial tachyon condensate phase of string cosmology
is followed by a period of cosmological inflation, then
the spectrum of fluctuations produced in the initial phase
will be observable via the exponential redshift of the
wavelength which the perturbation modes undergo. Thus,
the initial string phase of cosmology will set the initial
conditions for the fluctuation modes in the ultraviolet regime
at the beginning of the period of inflation. 

In this way, our study connects to  
one of the conceptual problems \cite{RHBrev0} of inflationary
cosmology, namely the \textit{trans-Planckian problem}. The 
issue is the following: provided that
the period of inflation lasted more than 70 e-foldings, then
the physical wavelengths of all fluctuations which are currently
observed have a physical wavelength at the beginning of the
period of inflation which is smaller than the Planck scale.
Hence, as first studied explicitly in \cite{Jerome}, the
predictions of the theory for observations are sensitive to
the unknown ultraviolet physics which determines the evolution
of fluctuations at early times in the ultraviolet (UV) region. 
In the context of scalar field-driven inflation, there have been several
approaches to deal with this problem - making use of dispersion
relations which are modified in the UV region \cite{Jerome,moddisp},
incorporating the effects of space-space \cite{spacespace} or
space-time \cite{spacetime} non-commutativity, appealing to a
minimal length \cite{Kempf}, assuming a ``new physics hypersurface''
\cite{screen}, or making use of effective field theory
techniques \cite{EFT} (see e.g. \cite{TPrev} for a review and
for more references).\footnote{Back reaction constraints on the 
magnitude of trans-Planckian
effects have been investigated in \cite{BR}.}

The key point is that, since the cosmological
fluctuations simply red-shift in an expanding universe, the imprints
of trans-Planckian physics are not diluted but simply
red-shift. Thus, imprints of trans-Planckian physics are, in
principle, measurable in current observations. The \textit{trans-Planckian
problem} for inflationary cosmology has become the \textit{trans-Planckian
window of opportunity} to probe Planck-scale physics today.
To realize this window of opportunity, however, it is
crucial to correctly couple the cosmological fluctuations to the
fundamental theory which describes UV physics. 

In this paper, we will address this question in the context of a recent
perturbative approach to singularity resolution in string theory
via the condensation of closed string tachyons \cite{tachyoncond}.
To be specific, we will consider the setup of \cite{Eva} in which
the condensation of closed string winding modes is considered
in a contracting homogeneous universe. Once the radius of space
descreases below a critical value, the winding modes become
tachyonic, and a tachyon condensate forms. As the tachyon
condensate increases in magnitude, all string fluctuations, including
those which correspond to cosmological perturbations, are frozen
out. One interpretation of this phenomenon is that the dimensions
of space in which the tachyon condenses disappear, and that thus
a topology change occurs (see e.g. \cite{topchange} for further
references on this interpretation). Here, we will consider the
time reverse of this dynamics: our universe is emerging from a
tachyonic big bang. As the tachyon condensate decreases, fluctuation
modes gradually emerge. We will track these fluctuations from
the time when they emerge until the time $t_0$ when the tachyon condensate
has disappeared, and we will compare the result with what would
be obtained from naive vacuum initial conditions of the low
energy field theory fluctuations at the time $t_0$. 

We find that in our setup, cosmological fluctuations emerge in
a thermal state. If the early tachyonic phase of cosmology connects
to the late time universe via a phase of inflation, and 
the period of inflation is comparable to the minimal period required
for inflation to solve the cosmological puzzles of standard big bang
cosmology, then we obtain a thermal rather than a scale-invariant
spectrum of fluctuations. Only if inflation lasts much
longer than the minimal number of e-foldings, then 
scales being probed in observations today
would correspond to the far Wien tail of the thermal spectrum, and thus
the predictions for observations would be similar to what would be
obtained assuming the usual vacuum state for cosmological perturbations.
If the initial tachyon phase connects to our present universe via
the recently suggested string gas structure formation scenario,
the tachyonic phase studied in this paper could provide the
required thermal initial conditions.

The outline of this paper is as follows: in the next section
we will derive the equations of motion for cosmological 
perturbations which hold given a period of tachyon condensation.
In Section 3, we will solve these equations and compute,
making use of the Bogoliubov mode mixing technique, the
spectrum of perturbations at the end of the tachyon phase given
an initial ground state. We then discuss the implications of
our results for cosmology and conclude with a summary section.

\section{Equations of Motion for the Fluctuations}

\begin{figure}
\includegraphics[height=6cm]{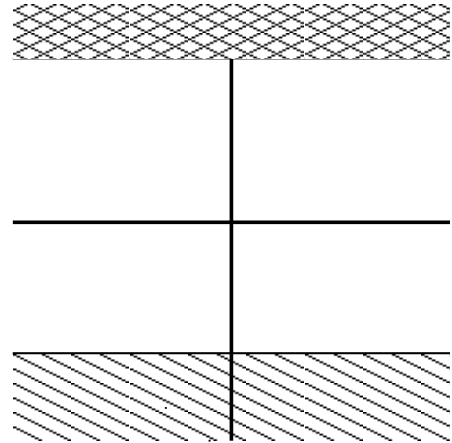}
\caption{Space-time sketch of our tachyon condensate background.
The vertical axis denotes time, the horizontal axis is space.
The bottom dashed region indicates the region of strong
tachyon condensation when it does not make sense to talk
about space. The upper cross-hatched region is the region of strong
coupling due to the linear dilaton, which we assume is stabilized to
yield a standard cosmology. Our
analysis studies the emergence of fluctuations in the
intermediate un-hatched weak tachyon phase.} \label{fig:1}
\end{figure}

In this section we will derive the equation of motion
for cosmological fluctuations which hold in the weak tachyon
period of tachyon condensation (see figure \ref{fig:1}).  
Our approach will be to 
consider two different string backgrounds as toy model cosmologies
and find that the equations of 
motion for fluctuations share the same features.  
We will work in the string frame and in tree-level (flat worldsheet)
string perturbation theory throughout.

\subsection{Tachyon Condensation and Dimension Change}

Even though we imagine a cosmology such as that of \cite{Eva} in which
the closed string tachyon is localized to a region of spacetime with a 
small circle, we will find it simpler to consider models in which the 
tachyon is present throughout the entire spacetime, as in \cite{topchange}.
In this case, the spacetime metric is flat at 
all times, but the dimensionality of spacetime is reduced in regions of
large tachyon.  Therefore, some (or all)
of the spacetime is in a noncritical dimension, so there must be a dilaton
gradient.  We will choose the dilaton gradient to be timelike in the 
region of smaller tachyon (and larger dimensionality), and we will proceed 
to ignore the dilaton gradient because it will modify the equations of motion
only by a ``Hubble-damping/forcing'' type term, and we prefer to focus
on other aspects of the physics.

From the point of view of the 2D worldsheet theory, 
the dimensions of spacetime are just
quantum fields, and the tachyon condensate induces a (spacetime-dependent)
mass for some of these fields.  Therefore, 
the worldsheet field theory flows to a target spacetime of lower dimension
as the massive fields are integrated out; 
most work has been concerned with
arguing for that RG flow \cite{Eva,topchange}.  (Note that this argument
requires that the tachyon itself depend on the spatial dimensions which it
eliminates, or else those dimensions would remain massless as worldsheet 
fields.)  On the other hand, we are
interested in the spacetime emerging from the tachyon condensate, so 
we will study the conformal interacting worldsheet theory relevant
for regions of spacetime with a small tachyon.  To keep matters simple, we
will work with the bosonic string (or equivalently the bosonic half of a 
nonsupersymmetric heterotic string).  Additional details and further results
on this theory from the perspective of the small tachyon region will be
presented in \cite{arf}.

\subsection{Backgrounds and Gauges}

As we mentioned above, the spacetime in the string backgrounds we consider
are flat (in the string frame), and we imagine the tachyon to have 
a spacetime profile of the form
\eq{timelike}{
T(X^\mu) \propto e^{-2\beta X^0} \vec X^2}
or 
\eq{null}{
T(X^\mu) \propto e^{-2\beta X^+} \vec X^2\ .}
Here, $\vec X$ runs over some of the spatial dimensions (which we
take for simplicity to be all the transverse 
spatial dimensions), which are the spatial
dimensions destroyed in the region of large tachyon condensate, and $\beta$
controls the gradient (and is generally related to the 
dilaton gradient).  These profiles are appropriate for the bosonic string
theory (up to the addition of an additional space-independent term which is
unimportant for our purposes); 
to get the same worldsheet action for the heterotic string, we
take $T\to T^2$ in the profiles (\ref{timelike},\ref{null}).  
Specifically, in either case, the worldsheet action for the string 
(ignoring the dilaton) will be 
\eq{action}{
S=-\frac{1}{4\pi\alp}\int_0^\ell d\tau d\sigma
\left[ \del_a X^\mu \del^a X_\mu
+\frac{\mu^2}{2} e^{-2\beta t} \vec X^2\right] ,
}
where $t=X^0$ or $t=X^+$ respectively and $\ell$ is the coordinate length
of $\sigma$ on the worldsheet (with periodic boundary conditions).  
Note that the tachyon background
does introduce other terms in the action (depending only on $t$ in the
bosonic string and involving fermions in the heterotic string), but these
will not affect our analysis of the bosonic modes.

From action (\ref{action}), it seems that the obvious gauge choice would be
to take $t=\alp \hat p\tau$, relating spacetime and worldsheet times
(where $\hat p=p^{0,+}$ for shorthand).  To get to
this gauge, we first take worldsheet coordinates such that the worldsheet
metric is conformally flat.  In these coordinates, $t$ satisfies a wave
equation on the worldsheet, so we can write $t=f(\bar\tau+\bar\sigma)
+g(\bar\tau-\bar\sigma)$.
Then we can take $\tau=t$ and $\sigma=f(\bar\tau+\bar\sigma)
-g(\bar\tau-\bar\sigma)$, which leaves the metric conformally flat.  Because
the theory is conformally invariant, we can just use the flat worldsheet 
metric.  As a final note, we can set the worldsheet spatial coordinate
periodicity to $\ell=2\pi$ 
by concurrently rescaling the tachyon amplitude $\mu$
(or equivalently shifting $X^{0,+}$).

There are,
however, caveats in the two cases.  First, for the timelike tachyon,
this gauge is only sensible if we treat the tachyon contribution to the
action as a perturbation and set the zeroth order part of $t$ proportional
to $\tau$.  This follows because higher order parts of $X^0$ are not
harmonic on the worldsheet according to their equations of motion.  
Also, in order to avoid $\alp$ corrections to the tree-level quantization,
we need small $\beta$ and therefore a large positive dilaton gradient.  This 
means that string theory becomes strongly coupled in the future.  We will
deal with this objection by supposing the strong coupling region to be
in the far future, after the tachyon generates fluctuations.  In a more
realistic model, we would also expect the dilaton to stabilize at some finite value.

For the null tachyon gradient, we no longer have to worry about $\alp$ 
corrections, as discussed in \cite{topchange}.  However, while we have
chosen a lightcone gauge, the dilaton gradient is still timelike in a 
supercritical number of dimensions.  To circumvent this
difficulty, we choose to have critical dimension in the weak tachyon region.  
In this case, the tachyon profile (\ref{null}) solves the equations of motion
for a lightlike dilaton $\Phi\propto X^-$, which does not interfere with the
lightcone quantization. Note that 
there is again a strongly-coupled region of spacetime (both due to the 
lightlike dilaton in the weak tachyon region and a spacelike dilaton gradient
in the strong tachyon region).  
However, again, we can
push the strong-coupling region arbitrarily far away, so we expect that
it does not affect our results.

\subsection{Worldsheet Hamiltonian}

To quantize the string, we first need to work out the mode expansion of
the transverse coordinates $\vec X$.  To first order in the tachyon background,
we find
\bea
\vec X &=& \vec x \left(1-\frac{1}{4} \mbp e^{-2\alp\beta\hat p\tau}\right)
\nonumber\\
&+&\alp \vec p \tau\left(1-\frac{1}{2}\mbp
(1+\beta t)e^{-2\alp\beta \hat p\tau}\right)\nonumber\\
&+&i\sqrt{\frac{\alp}{2}}\sum_{n\neq 0} \frac{1}{n}\left[\vec{\tilde\alpha}_n
e^{-in(\tau+\sigma)} \left( 1\frac{}{}
\right.\right.\nonumber\\
&-&\left.\left.\frac{1}{4(1+\nu)}\mbp e^{-2\alp\beta \hat p\tau}\right)+
\vec\alpha_n e^{-in(\tau-\sigma)}\right.\nonumber\\
&&\left.\times \left(1-\frac{1}{4(1-\nu)}
\mbp e^{-2\alp\beta \hat p\tau}\right)\right] \,\label{modes}
\eea
where $\nu=i n/\alp\beta\hat p$.
This is the usual mode expansion with an exponentially dying correction for
each mode (including the center of mass coordinate).  To avoid working with
worldsheet fermions, we specialize to the bosonic string.

There are two ways to derive (\ref{modes}).  One is to note that the
equation of motion
\eq{bessel1}{
\left(-\del_\tau^2 + \del_\sigma^2 \right)\vec X = \mu^2 
e^{-2\alp\beta\hat p\tau} \vec X}
has for each spatial mode $\exp(in\sigma)$ the solution
\eq{bessel2}{
\vec X_n \!=\! \vec a_n J_\nu \left(\frac{\mu}{\alp\beta\hat p}
e^{-\alp\beta\hat p\tau}\right)\!\!+\vec b_n Y_\nu \left(\frac{\mu}{\alp\beta
\hat p}e^{-\alp\beta\hat p\tau}\right)\!.}
Then we can simply expand around late times and match the leading terms
to the usual mode expansion.  Alternatively, we can split the equation of 
motion (\ref{bessel1}) and oscillator $\vec X$ into zeroth and first order
parts, which gives exactly the correction terms in (\ref{modes}) as the
first order part of (\ref{bessel2}).

From the mode expansion (\ref{modes}), we can write out the worldsheet
Hamiltonian in our two gauges as
\bea
H&=& \frac{1}{2} 
\alp \vec p^2 \left(1+\mbp (1 +\beta t)^2
e^{-2\beta t}\right)\nonumber\\
&&+\sum_{n=1}^\infty \left[(\vec\alpha_{-n}\cdot\vec\alpha_n +
\vec{\tilde\alpha}_{-n}\cdot\vec{\tilde\alpha}_n)
\left(1\vphantom{\frac{1}{n^2}}\right.\right.\nonumber\\
&&\left.\left.+\frac{\mu^2}{4n^2}e^{-2\beta t}\right)\right]
- 2 +\left\{\frac{\alp \hat p^2}{2}\right\}
+\cdots\ ,
\label{ham1}
\eea
in which the term in curly braces only appears for the static gauge case.
Also, in the static gauge case, we need to add
the usual momentum term without corrections 
for the longitudinal direction (usually $X^1$), but
not the oscillators (because they are not independent degrees of freedom).
We have denoted by ellipses terms involving the center of mass position
and terms which change the state of the string.  For our first analysis,
we will ignore these effects.  
Finally, there are terms correcting the worldsheet
zero-point energy.  
Considering the exact results of \cite{topchange} in the lightlike tachyon
case, we know that these corrections to the zero-point energy must vanish
once the linear dilaton and all quantum effects have been taken into 
account.

\subsection{Model Equation of Motion}

To relate the worldsheet Hamiltonian to the physics of fluctuations in
spacetime, consider the worldsheet Schr\"odinger equation
\eq{1final}{
H=i\del_\tau = i \alp\hat p\del_t \ .}
This last derivative is just the momentum conjugate to $t$, either $ip_0$
or $ip_+$ in the timelike and lightlike case respectively.
Then we have
\eq{ham2}{
H= \alp\times\left\{\begin{array}{c}
\left(p^0\right)^2 \\\textnormal{or} \\ p^+ p^-\end{array}\right.\ .}

Comparing (\ref{ham1},\ref{ham2}), we can write the Schr\"odinger equation
as a mass-shell relation,
\bea
0&=& -(p^0)^2+ \vec p^2 \left(1+\left(\frac{\mu}{\alp\beta p^0}\right)^2 
(1 +\beta t)^2 e^{-2\beta t}\right)\nonumber\\
&&+\frac{\mu^2}{2\alp}e^{-2\beta t}\label{mass1}
\eea
or
\bea
0&=& -2p^+p^- + \vec p^2 \left(1+\left(\frac{\mu}{\alp\beta p^+}\right)^2 
(1 +\beta t)^2 e^{-2\beta t}\right)
\nonumber\\
&&+ \frac{\mu^2}{2\alp} e^{-2\beta t}\ .\label{mass2}\eea
We have specialized here to the case of massless string modes (such as
the dilaton and graviton), which have 
$\langle\vec\alpha_{-1}\cdot\vec\alpha_1\rangle=
\langle\tilde{\vec\alpha}_{-1}\cdot\tilde{\vec\alpha}_1\rangle =1$ and all
other oscillators unexcited.

It is worth commenting on the corrections to the Minkowski mass-shell 
relation $p^2=0$.  The simplest term is the common last term of 
(\ref{mass1},\ref{mass2}), which is just a time-dependent mass term for the
string mode.  This term arises because the $\vec X$ oscillators have a 
time-dependent excitation energy on the worldsheet.  From a spacetime point
of view, this mass term must arise from a graviton-tachyon coupling.  
The other correction visible in (\ref{mass1},\ref{mass2}) is a modified
dispersion relation, which essentially corresponds to a time-varying
speed of light.  Some points need to be made about this term.  First,
the time-dependence is no longer a pure exponential.  However, the correction
is subleading, so we will ignore it for simplicity.  Second, the coefficient
of the correction contains either $1/(p^0)^2$ or $1/(p^+)^2$ in the two 
cases.  This term appears problematic because the momenta should be 
represented as derivatives in the equation of motion.  However, we will
treat them (in this correction term only) as constants, for the following
reason.  In the null tachyon case, $p^+$ commutes with $t$ and is in
fact a constant of motion.  Similarly, $p^0$ is a constant of motion for
the timelike tachyon background at zeroth order in the tachyon.  Nonetheless,
this is a curious dependence, and we will return to this issue in 
\cite{arf}.

We close this section with a few comments on the terms we have neglected.
First, there are explicitly position-dependent terms in the mass-shell
relations.  This means that momentum eigenmodes are not actually eigenmodes
of the full system; we expect this result because the tachyon varies through
space.  Next, there are also terms mixing different oscillation states of
the string, so the normal Fock space of string excitations is not an 
eigenbasis.  Essentially, the time-varying tachyon allows the string to 
grow or shrink in time.

\section{Solutions of the Equations}

\subsection{Normalized Form of the Equation of Motion}

In this section we will discuss approximate solutions of
the equation derived in the previous section, focusing on the time-like case
due to its relevance to cosmology (the light-like tachyon gives us
confidence in our results, however).
The starting point is the mass-shell condition (\ref{mass1}), which 
is also the equation of motion for the 
linearized cosmological
perturbation mode $\psi$ in the presence of a time-like tachyon
condensate. The Fourier mode of $\psi$ with comoving wave number $k$
satisfies the equation
\eq{wf1}{
\left[\frac{\partial^2}{\partial t^2}
+\left(c_2\frac{\mu^2}{\alp}+c_1k^2\right)e^{-2\beta t}
+k^2\right]\psi(t)=0\,.
}
Here, $c_1$ and $c_2$ are coefficients determining the strength of the 
corrections to the mass and dispersion relation.

This equation describes the evolution of a harmonic oscillator
with a mass term which depends on the tachyon condensate and
thus depends exponentially on time. At very early times
$t \rightarrow - \infty$, the frequency of oscillation grows
exponentially, whereas in the late time limit $t \rightarrow + \infty$,
$\psi$ is oscillating with its bare mass. As a consequence of the
time dependence of the mass we expect particle production. A
mode which is pure positive frequency in the past will be
composed of a mixture of positive and negative frequency modes
in the future. The mode mixing is given by the Bogoliubov mode
mixing matrix.

If we define the following variables
\eq{defs}{
-2\beta t\equiv X
\ ,\ \
\frac{1}{4\beta^2}\left(c_2\frac{\mu^2}{\alp}+c_1k^2\right)\equiv\lambda
\ ,\ \ 
\frac{k^2}{4\beta^2}\equiv\omega^2\,,
}
then Eq.~(\ref{wf1}) can be rewritten in the normalized form
\begin{equation}
\left[\frac{\partial^2}{\partial X^2}
+\lambda e^X+\omega^2\right]\psi(X)=0
\label{wf2}\,.
\end{equation}
This is the same equation which was studied by Strominger \cite{Andy} and
collaborators \cite{GS} in their analysis of open string
production by decaying S-branes. To make the correspondence with
the equation of motion studied in \cite{GS}, we must make the
substitution  
\begin{equation}
\omega^2=N-1+k^2\,.
\end{equation}
In contrast to the analysis in \cite{GS}, in our case 
we have $k$ dependence in $\lambda$. Thus,
we can expect that the Bogoliubov coefficients which describe
string production in \cite{GS} and the production of
cosmological perturbations in our case will have a 
different $k$ dependence.

\subsection{Bogoliubov Coefficients}

Let us now calculate the Bogoliubov coefficients in this setup.
To describe the initial vacuum state, we must consider the mode
which is correctly normalized and pure positive frequency in
the past and then expand this mode in the asymptotic future
in terms of correctly normalized positive frequency and negative
fequency states. The expansion coefficients are the Bogoliubov
coefficients we desire. Since the Bogoliubov mode mixing matrix is 
unitary, we can equivalently take the vacuum state at future
infinity and expand it in terms of the correctly normalized
positive and negative frequency modes in the far past. This is
what will be done below.
 
Since Eq.~(\ref{wf2}) can be rewritten by using the variable 
$y=2\sqrt{\lambda}e^{X/2}$:
\begin{equation}
\left[\frac{\partial^2}{\partial y^2}
+\frac{1}{y}\frac{\partial}{\partial y}+1
+\frac{4\omega^2}{y^2}\right]\psi(y)=0
\label{wf3}\,,
\end{equation}
we see that the solutions are Bessel functions: 
$\psi\propto J_{\pm 2i\omega}(y)$, more precisely,
\begin{equation}
\psi^{\mathrm{out}}_{\vec{k}}
=\frac{\lambda^{i\omega}}{\sqrt{2\omega}}\Gamma(1-2i\omega)
e^{i{\vec{k}\cdot\vec{x}}}J_{-2i\omega}(2\sqrt{\lambda}e^{\frac{X}{2}})\,,
\label{solution}
\end{equation}
where we chose the above particular solution to be the
canonically normalized positive 
frequency mode in the far future $X\rightarrow-\infty$ (which
describes the vacuum state in the future)
\begin{eqnarray}
&& \frac{\lambda^{i\omega}}{\sqrt{2\omega}}\,\Gamma(1-2i\omega)
\,e^{i{\vec{k}\cdot\vec{x}}}\,J_{-2i\omega}(2\sqrt{\lambda}e^{\frac{X}{2}})
\, \nonumber \\
&\cong& \,
\frac{1}{\sqrt{2\omega}}e^{-i\omega X+i{\vec{k}\cdot\vec{x}}} \, ,
\end{eqnarray}
where we have used the well-known asymptotic form
\be
J_\nu(z)
\cong\left(\frac{z}{2}\right)^\nu\frac{1}{\Gamma(\nu+1)} \quad
{\rm for} \quad z\rightarrow 0
\ee
of the Bessel functions.

On the other hand, in the far past $X\rightarrow\infty$, the 
solution (\ref{solution}) becomes
\begin{eqnarray} \label{in}
\psi^{\mathrm{out}}_{\vec{k}}\, &\rightarrow& \,
\frac{\lambda^{i\omega-1/4}}{\sqrt{8\pi\omega}}\,\Gamma(1-2i\omega)\,
e^{-\frac{X}{4}+i{\vec{k}\cdot{\vec{x}}}}\, \\
&& \left[\,e^{\pi\omega-2i\sqrt{\lambda}e^{X/2}+\frac{\pi}{4}i}
+e^{-\pi\omega+2i\sqrt{\lambda}e^{X/2}-\frac{\pi}{4}i}\,\right]
\nonumber
\end{eqnarray}
where we have made use of the asymptotic formula for $z\rightarrow \infty$
\bea
J_\nu(z) &\cong& \sqrt{\frac{2}{\pi z}}\cos(z-\frac{2\nu+1}{4}\pi) \\
&=&  \frac{1}{\sqrt{2\pi z}}\left[\,e^{iz-i\frac{\pi}{2}\nu-\frac{\pi}{4}i}
+e^{-iz+i\frac{\pi}{2}\nu+\frac{\pi}{4}i}\,\right] \nonumber
\eea

We thus find that the outgoing modes $\psi^{\mathrm{out}}_{\vec{k}}$ contain
both negative and positive frequency parts in the far past. This indicates
particle creation.

Next we look for the correctly normalized positive frequency modes in
the past ($y \rightarrow \infty$). 
Making the field redefinition $\psi=\sqrt{2/y}\,f$ in 
Eq.~(\ref{wf3}), we have
\begin{eqnarray}
\frac{\partial^2f}{\partial y^2}+f+O\left(\frac{1}{y^2}\right)f=0
\end{eqnarray}
Since in the far past the last term on the right-hand side is
neglibile, the early time vacuum modes (the correctly normalized 
positive frequency modes) are  
\be
f=\frac{1}{\sqrt{2}}e^{-iy}
\ee
which correspond to
\be
\psi^{\mathrm{in}}_{\vec k}=\frac{\lambda^{-1/4}}{\sqrt{2}}
e^{-\frac{X}{4}-2i\sqrt{\lambda}e^{X/2}+i\vec{k}\cdot\vec{x}}\,.
\label{out}
\ee

The corresponding particular solution of the full equation which
reduces to the above in the asymptotic limit is
\be
\psi^{\mathrm{in}}_{\vec k}=\sqrt{\frac{\pi}{2i}}\,
e^{{-\pi\omega}+i\vec{k}\cdot\vec{x}}\,
H^{(2)}_{-2i\omega}(2\sqrt{\lambda}e^{\frac{X}{2}})\, ,
\ee
where we have made use of the asymptotic formula
\be
H^{(2)}_{\nu}(z)\sim\sqrt{\frac{2}{\pi z}}e^{-iz+i\frac{\pi}{4}(2\nu+1)}
\, .
\ee

Comparing Eqs.~(\ref{in}) with (\ref{out}), we can find the 
Bogoliubov mode mixing relations
\be
\psi^{\mathrm{in}}_{\vec{k}}\,=\,\alpha_{\vec{k}}\,
\psi^{\mathrm{out}}_{\vec{k}}
+\beta_{\vec{k}}\,\left(\psi^{\mathrm{out}}_{-\vec{k}}\right)^*\,,
\ \ \ 
X\rightarrow\infty
\label{outin}
\ee
where
\begin{eqnarray}
\alpha_{\vec{k}}
&=&\frac{\lambda^{-i\omega}}{\sqrt{4\pi i\omega}}\,\Gamma(1+2i\omega)\,
e^{\pi\omega}\\
\beta_{\vec{k}}&=&
-\frac{\lambda^{i\omega}}{\sqrt{4\pi i\omega}}\,\Gamma(1-2i\omega)\,
e^{-\pi\omega}\ .
\end{eqnarray}
Here we used the relation
\be
\Gamma(1+2i\omega)\,\Gamma(1-2i\omega)=\frac{2\pi\omega}{\sinh{2\pi\omega}}
\, .
\ee
It is easy to check that the unitarity relation
\be 
\alpha_{\vec{k}}\alpha_{\vec{k}}^*-\beta_{\vec{k}}\beta_{\vec{k}}^*=1
\ee
is satisfied.

Note that in the case of our current analysis, we have a
$k$ dependence in $\lambda$, and thus the Bogoliubov coefficients 
have a different $k$-dependence than that is obtained in \cite{GS}. 

Incidentally, there are solutions to the equation of motion (\ref{wf3})
with $\omega^2<0$; these are just the same Bessel functions as in 
(\ref{solution}) but with (real) even integer order \cite{Fredenhagen:2003ut}.
Only the Bessel functions of the first kind are normalizable, so only 
they are allowed.  These die off as $t\to\infty$, so they do not affect our
analysis of Bogoliubov coefficients.  Also, these modes have imaginary 
spatial momentum $k$, and the normalizable ones are localized near the 
origin.  Physically, these modes represent localized transient massless string
excitations.

\subsection{Planck Distribution}

Let us now calculate the spectrum of the produced particles. First,
we expand the field operator $\hat{\psi}$ in terms
of creation and annihilation operators $a^{\dagger}$ and $a$
associated with the positive and negative frequency modes in both the
asymptotic past and future:
\be
\hat{\psi}=a_{\mathrm{out}}\psi^{\mathrm{out}}+a_{\mathrm{out}}^\dagger
\left(\psi^{\mathrm{out}}\right)^*
=a_{\mathrm{in}}\psi^{\mathrm{in}}+a_{\mathrm{in}}^\dagger
\left(\psi^{\mathrm{in}}\right)^*\,.
\ee
Substituting $\psi^{\mathrm{in}}$ by Eq.~(\ref{outin}), we get
\begin{eqnarray}
\alpha_{\vec{k}}\,\,a_{\mathrm{in},\vec{k}}
+\beta^*_{\vec{k}}\,\,a^\dagger_{\mathrm{in},-\vec{k}}
\,&=&\,a_{\mathrm{out}}\\
\beta_{\vec{k}}\,\,a_{\mathrm{in},-\vec{k}}
+\alpha^*_{\vec{k}}\,\,a^\dagger_{\mathrm{in},\vec{k}}
\,&=&\,a^{\dagger}_{\mathrm{out}}\,.
\end{eqnarray}

If we assume the vacuum $a^{\mathrm{in}}|0\rangle=0$ at $X\rightarrow \infty$,
the number of strings at $X\rightarrow -\infty$ is expressed as
\begin{eqnarray}
\langle0|a^\dagger_{\mathrm{out}}a_{\mathrm{out}}|0\rangle
&=&|\beta|^2\langle0|a_{\mathrm{in}}a_{\mathrm{in}}^\dagger |0\rangle
=|\beta|^2\nonumber\\
&=&\frac{1}{4\pi\omega}\Gamma(1-2i\omega)\Gamma(1+2i\omega)e^{-2\pi\omega}
\nonumber\\
&=&\frac{1}{e^{4\pi\omega}-1}=\frac{1}{e^{\frac{2\pi}{\beta}k}-1}
\end{eqnarray}
Thus we see that the resulting spectrum has a thermal distribution with
a temperature given by $\beta/(2\pi)$.

In conclusion, we have shown that cosmological fluctuation modes
emerge from the tachyon condensate phase in a thermal state with
temperature given by $\beta/(2 \pi)$. This result is analogous to
what happens in the context of string production by S-branes. The
thermality is related to the fact that the tachyon condensate is
periodic in imaginary time \cite{GS}.

\section{Connections to Cosmology}

We have found that the spectrum of cosmological perturbations which
emerges from a tachyonic big bang is a thermal spectrum rather than
a vacuum. The way in which this result relates to cosmological
observations obviously depends on how the initial tachyonic phase of
string cosmology is connected to late time cosmology. It is usually
assumed that the connection occurs through a period of cosmological
inflation (see e.g. \cite{stringinflation} for recent
reviews on how inflation might arise from string theory). However, 
one must keep in mind that there are alternative mechanisms to create a 
spectrum of coherent, almost-adiabatic, nearly Gaussian, and almost 
scale-invariant cosmological perturbations (such as those
observed).  Such spectra were postulated long before inflation 
\cite{Harrison,Zel,SZ}. Inflationary cosmology \cite{Guth}
provided the first successful model which predicted such a spectrum,
but recently other ideas have been proposed. In particular,
it was suggested that the radiation phase of
standard cosmology is preceded by a quasi-static initial Hagedorn 
phase \cite{BV} of perturbative string theory, and the
thermal string fluctuations in this phase then lead to a
spectrum of cosmological perturbations with the desired
properties \cite{NBV,BNPV2} (see \cite{SGfluctsrev} for a
recent brief review).

Let us first consider the implications of our study for the case
of cosmological inflation. In this case, the modes which are
probed in current cosmological observations emerged deep in
the ultraviolet sea at the beginning of inflation. Our analysis
provides the initial conditions for these modes, and according
to our analysis the initial conditions differ substantially
from the Bunch-Davies vacuum initial conditions which are usually
assumed. 

Let us denote the Hubble parameter at the beginning of the
period of inflation by $H_I$. If inflation is preceded by
a radiation phase (as is the case in our setup), then the
temperature of the radiation at the onset of inflation is\
given by
\be
T_I \, = \, \left( H_I m_{pl} \right)^{1/2} \, ,
\ee
where $m_{pl}$ is the reduced Planck mass. The
wavelength of a mode increases (redshifts) by a factor of
\be
F \, = \,  \tilde\beta \left( m_{pl} / T_I \right)
\ee
between the end of the tachyon phase, when the temperature
is 
\be
T_T \, = \, \tilde{\beta} m_{pl}
\ee
(here we write the tachyon condensate
gradient $\beta$ of the previous sections as a dimensionless
constant $\tilde{\beta}$ times the reduced Planck mass).
Hence, the wavelength $\lambda_{p}$ corresponding to the
peak of the thermal distribution has a length 
\be
\lambda_p(T_I) \, = \, \left( \frac{H_I}{m_{pl}} \right)^{1/2} H_I^{-1}
\ee
at the beginning of the period of inflation, which is much shorter than
the inflationary horizon length. 

If the period of inflation has the minimal length for inflation to
solve the horizon and flatness problems of standard cosmology, then
between the onset of inflation and the current time the wavelength
corresponding to the Hubble radius $H_I^{-1}$ at the beginning of
inflation redshifts to the length of the current Hubble radius.
In this case, given our tachyon condensate precursor phase to
inflation, we predict that the spectrum of cosmological fluctuations
should be thermal rather than scale-invariant. Inflationary models
with a much larger number of e-foldings of inflation, 
in which the initial inflationary Hubble radius redshifts to much more
than
$(m_{pl} / H_I)^{1/2}$ 
times the current Hubble radius by the present
time, would be safe in the sense that the specific trans-Planckian
signatures would have been redshifted beyond the realm of measurability
of observations. Observations would only detect the far Wien tail of the
thermal spectrum, which would appear scale-invariant.

In the case when our tachyonic big bang phase is the precursor
phase of the quasi-static Hagedorn phase, then our study would
support taking the initial conditions for fluctuations to be 
thermal, as assumed in \cite{NBV}. In fact, it is interesting
to speculate that a tachyonic big bang phase in fact is a
description of the stringy Hagedorn phase. A toy model along
these lines is explored in a parallel paper \cite{BFK2}.

\section{Conclusions}
 
In this paper, we have studied the emergence of cosmological
fluctuations from a tachyonic big bang phase. The result
of our approximate treatment is that the cosmological fluctuations
corresponding to zero mass sting excitations like the graviton
emerge in a thermal state, with a temperature given by the
tachyon profile and expected to be given by the string scale.

If the initial tachyon phase is connected to late time cosmology
via a period of cosmological inflation, then our work is
relevant to the question of possible trans-Planckian
signatures of inflation. In fact, in models of inflation with
a number of e-foldings close to the minimal number required for
inflation to solve the problems of standard cosmology, we
obtain a thermal rather than a scale-invariant spectrum.
In models in which inflation lasts much longer, current observations
would only probe the far Wien tail of the thermal spectrum, and our
analysis would then lend support to the use of vacuum initial conditions for
the fluctuating modes. 

However, our tachyon phase can also be
seen as providing the thermal initial conditions required by
the recently suggested string gas cosmology structure formation
scenario.

\begin{acknowledgments} 

We wish to thank Gary Horowitz, Alex Maloney, Maulik Parikh,
Eva Silverstein, Jiro Soda and in particular Simeon Hellerman for 
fruitful discussions. One of us (RB) acknowledges
hospitality and financial support by the KITP in Santa Barbara during
the mini-program on the {\it Quantum Nature of Spacetime Singularities}
in January 2007 during which the idea for this project arose.
AF is supported in part by a Institute for Particle Physics postdoctoral
fellowship, and SK is supported by a grant for research abroad by the
JSPS. The work of RB and AF by an 
NSERC Discovery Grant and by the Canada Research Chairs program. 
 
\end{acknowledgments}

\end{document}